\pdfoutput=1

\documentclass[aps,prl,10pt,twocolumn,showpacs,superscriptaddress,footnoteinbib]{revtex4}
\usepackage{amsmath}
\usepackage{latexsym}
\usepackage{amssymb}
\usepackage{graphics,epstopdf}

\begin{document}

\title{Fine-grained uncertainty relation and biased non-local games in bipartite and tripartite systems}

\author{Ansuman Dey}
\thanks{ansuman@bose.res.in}

\author{T. Pramanik}
\thanks{tanu.pram99@bose.res.in} 

\author{A. S. Majumdar}
\thanks{archan@bose.res.in}
\affiliation{S. N. Bose National Centre for Basic Sciences, Salt Lake, Kolkata 700 098, India}


\begin{abstract}
 
The  fine-grained uncertainty relation can be used to discriminate among 
classical, quantum and super-quantum correlations based on their strength of
non-locality, as has been shown for bipartite and tripartite systems with 
unbiased measurement settings. Here we consider the situation when two and 
three parties, respectively, choose settings with bias for playing certain 
non-local games. We show analytically that while the fine-grained uncertainty
principle is still able to distinguish classical, quantum and super-quantum
correlations for biased settings corresponding to certain ranges of the 
biasing parameters, the above-mentioned discrimination is not manifested for
all biasing.

\end{abstract}


\maketitle

\section{Introduction}

Heisenberg uncertainty relation \cite{HUR} infers the restriction inherently 
imposed by quantum mechanics that we cannot simultaneously predict the 
measurement outcomes of two non-commuting observables with certainty. This 
uncertainty relation was generalized for any two arbitrary observables by 
Schr$\ddot{o}$dinger and Robertson \cite{GUR}. In quantum information theory, 
it is more convenient to use the uncertainty relation in terms of entropy 
in stead of standard deviation. A lot of effort has been devoted towards 
improving entropic uncertainty relations \cite{IBB,Deu,Sur}, especially  
in terms of their practical applicability in several information processing
scenarios such as quantum information locking and 
key generation \cite{lock,code,Berta}. Recently, a new fine-grained form of
the uncertainty relation has been proposed \cite{FUR1} which is able to
distinguish between uncertainties inherent in various possible measurement
outcomes, and is linked with the degree of non-locality of the underlying 
theory.

Though the use of entanglement in information processing tasks is
widely appreciated,  quantum correlations may
not be advantageous compared to classical ones in all types of situations.
Since entanglement is a fragile resource, the question as
to when precisely quantum non-locality following from entanglement is necessary
for practical applications, is  rather important. In this context, the
application of the fine-grained uncertainty relation could be particularly
relevant. The  fine-grained uncertainty relation is able to discriminate 
between the 
degree of non-locality in classical, quantum and super-quantum correlations
of bipartite systems, as was shown by Oppenheim and Wehner \cite{FUR1} in the 
context of a  class of non-local retrieval games for which there exists only 
one answer for any of the two parties to come up with in order to win. The 
maximum probability of winning the retrieval game is equal to the upper bound 
of the uncertainty relation and this quantifies the degree of non-locality of 
the underlying physical theory. This upper bound could thus be used to 
discriminate among the degree of non-locality pertaining to various underlying
theories such as classical theory, quantum theory and no-signaling theory with 
maximum non-locality for bipartite systems. 

Further insight into the nature of difference between various types of
correlations has been recently provided by the 
work of Lawson \textit{et al.}\cite{tlawson}. For a class of 
Bell-CHSH \cite{bell,bell2} games, they 
introduce the situation when the two parties decide to choose their 
measurements with bias. It has been shown that for certain range of the biasing 
parameters, quantum theory offers advantage and surprisingly for others, it 
does not provide a better result than classical mechanics. This leads towards 
the identification of situations when quantum entanglement is indeed essential 
for implementing a particular information processing task. A generalization
for multipartite systems is also performed in which numerical results
for the upper bound of the correlation function is presented when all 
parties measure with equal bias.

In this work we  investigate the connection between the fine-grained 
uncertainty relation and non-locality in the context of biased non-local games 
for first bipartite and then tripartite systems. Our motivation is to
utilize the fine-grained uncertainty relation in order to determine the
nonlocal resources necessary for implementing this particular information
processing task of winning a biased game played by two or three parties.
Here we make use of the formalism developed by Oppenheim and Wehner \cite{FUR1}
for bipartite systems, and its subsequent extension
to the case of tripartite systems \cite{tanu}. In case of bipartite systems, 
correlations are expressible in terms of Bell-CHSH \cite{bell,bell2} form 
without ambiguity and can be used efficiently for above mentioned task of 
discrimination between classical, quantum and super-quantum theories. The 
scenario for the tripartite case is however, a bit different as there is 
an inherent non-uniqueness regarding the 
choice of correlations proposed by Svetlichny \cite{Svet} and Mermin 
\cite{Mermin}. It has been shown \cite{tanu} that the Svetlichny-type 
correlations can discriminate among the classical, quantum and no-signaling 
theory using the fine-grained uncertainty principle, whereas the inequality 
extracted from the Mermin-type correlation is unable to perform the same task. 
In our present analysis we use the approach proposed by Bancal 
\textit{et al.}\cite{bancal} in order to calculate the upper bound of the 
Svetlichny function analytically in case of the biased tripartite game.
We are thus able to present without using numerical methods our results on
the ranges of 
biasing parameters when quantum correlation are beneficial for non-local tasks.
In what follows we will first present the description of biased nonlocal
games as provided by Lawson \textit{et al.}\cite{tlawson}, using the
terminology of fine-grained uncertainty relations \cite{FUR1,tanu}. In the
process, we will recount several results of Ref.\cite{tlawson} for
the bipartite game in the next
section. The utility of our approach in deriving new analytical results 
will be clear in the section on tripartite games.
 
\section{Fine-grained uncertainty and biased nonlocal games}

We begin with the description of the scheme of the game to be played within 
the bipartite system. The situation is such that the two parties, namely, 
Alice and Bob share a state $\rho_{AB}$ which is emitted and distributed by a 
source. Alice and Bob are spatially separated enough so that no signal can 
travel while experimenting. Alice performs either of her measurements $A_{0}$ 
and $A_{1}$ and Bob, either of $B_{0}$ and $B_{1}$ at a time. These measurements 
having the outcomes $+1$ and $-1$, can be chosen by Alice and Bob without 
depending on the choice made by the other. The CHSH inequality \cite{bell2} 
\begin{equation}
\frac{1}{4} [E(A_{0}B_{0})+ E(A_{0}B_{1})+E(A_{1}B_{0})-E(A_{1}B_{1})]\leq \frac{1}{2}
\end{equation}
holds for any local hidden variable model and can be violated when measurements 
are done on quantum particles prepared in entangled states. Here $E(A_{i}B_{j})$
 are the averages of the product of measurement outcomes of Alice and Bob with $i,j=0,1$. 
The above inequality refers to the scenario when the two parties have no bias 
towards choosing a particular measurement. 

In the following picture, describing
the biased game\cite{tlawson}, the 
intention of Alice is to choose $A_{0}$ with probability 
$p$($0\leqslant p\leqslant 1$) and $A_{1}$ with probability $(1-p)$. Bob 
intends to choose $B_{0}$ and $B_{1}$ with probabilities 
$q$($0\leqslant q\leqslant 1$) and $(1-q)$,  respectively. The measurements 
and their outcomes are coded into binary variables pertaining to  an 
input-output process. Alice and Bob have binary input variables $s$ and $t$, 
respectively, and output variables $a$ and $b$, respectively. Input $s$ takes 
the values $0$ and $1$ when Alice measures $A_{0}$ and $A_{1}$, respectively. 
Output $a$ takes the values $0$ and $1$ when Alice gets the measurement 
outcomes $+1$ and $-1$, respectively. The identifications are similar for 
Bob's variables $t$ and $b$. Now, the rule of the game is that Alice and Bob's 
particles win (as a team) if their inputs and outputs satisfy 
\begin{equation}
a\oplus b = s.t
\label{rule}
\end{equation}
where $\oplus$ denotes addition modulo $2$. Input questions $s$ and $t$ have 
the probability distribution $p(s,t)$ (for simplicity we take $p(s,t)= p(s)p(t)$ where $p(s=0)=p$, $p(s=1)= (1-p)$, $p(t=0)= q$ and $p(t=1)= (1-q)$ in 
our case). 

The fine-grained uncertainty relation \cite{FUR1} may be now invoked 
by noting that
for every setting $s$ and the corresponding outcome $a$ of Alice one may
formally denote a string  $\textbf{x}_{s,a}= (x_{s,a}^{1}, x_{s,a}^{2})$ determining the winning answer 
$b= x_{s,a}^{t}$($\forall x\in(0,1) $) for Bob; $\lbrace s\rbrace \in \mathcal{S}$ and $\lbrace t\rbrace \in \mathcal{T}$, $\mathcal{S}$ and $\mathcal{T}$ being the set of Alice's 
and Bob's input settings, respectively.  Alice and Bob receive the binary 
questions $s,t\in \lbrace 0,1\rbrace$ (i.e. representing two different 
measurement settings on each side) with corresponding probabilities (for $p$ 
and $q$ $\neq$ 0,1 the game is non-local) and they win if their respective 
outcomes $a,b\in \lbrace 0,1 \rbrace$ satisfy the condition ($\ref{rule}$). 
Before starting the game (a biased CHSH-game), Alice and Bob communicate and 
discuss their strategy, i.e., choice of the bipartite state $\rho_{AB}$ they 
are sharing and their measurements. They are not allowed to communicate once 
the game starts. The probability of winning the game for a physical theory 
described by bipartite state ($\rho_{AB}$) is given by \cite{FUR1},
\begin{equation}
P^{game}(\mathcal{S}, \mathcal{T},\rho_{AB})= \sum_{s,t} p(s,t)\sum_{a} p(a,b=x_{s,a}^{t}|s,t)_{\rho_{AB}}
\label{fgu1}
\end{equation}
 When $P^{game}(\mathcal{S}, \mathcal{T},\rho_{AB})$ is less than $1$, the 
outcome of the game is uncertain. The value of $P^{game}$ is bound by particular
 theories. For the unbiased case (i.e., $p(s,t)=p(s).p(t)=\frac{1}{2} .\frac{1}{2}= \frac{1}{4}$), the upper bounds of this value in classical, quantum and 
no-signaling theory are $\frac{3}{4}$, $\frac{1}{2}+\frac{1}{2\sqrt{2}}$ and 
$1$ respectively. The form of $p(a,b=x_{s,a}^{t}|s,t)_{\rho_{AB}}$ in terms of the 
measurements on the bipartite state $\rho_{AB}$ is given by,
\begin{equation}
p(a,b=x_{s,a}^{t}|s,t)_{\rho_{AB}}= \sum_{b} V(a,b|s,t)\langle (A_{s}^{a} \otimes B_{t}^{b})\rangle_{\rho_{AB}}
\label{fgu2}
\end{equation}
where, $A_{s}^{a}= \frac{\mathcal{I}+(-1)^{a}A_{s}}{2}$ is the measurement of the 
observable $A_{s}$ corresponding to the setting $s$ giving the outcome $a$ at 
Alice's side; $B_{t}^{b}= \frac{\mathcal{I}+(-1)^{b}B_{t}}{2}$ is a measurement 
of the observable $B_{t}$ corresponding to the setting $t$ giving the outcome 
$b$ at Bob's side and $V(a,b|s,t)$ filters the winning combination and is 
given by,
\begin{eqnarray}
V(a,b|s,t)&=& 1 ~~ \text{iff $a\oplus b=s.t$} \nonumber \\
&=&0 ~~\text{otherwise.}
\label{filter}
\end{eqnarray}
$P^{game}(\mathcal{S},\mathcal{T},\rho_{AB})$ can now be calculated using the 
Eqs.(\ref{fgu1})-(\ref{filter}) with the given probabilities of different 
measurements of Alice and Bob. For the bipartite state $\rho_{AB}$, the 
expression of $P^{game}$ is given by 
\begin{equation}
P^{game}(\mathcal{S},\mathcal{T},\rho_{AB})= \frac{1}{2}[1+ \langle CHSH(p,q)\rangle_{\rho_{AB}}]
\end{equation}
with $CHSH(p,q)= [pq A_{0}\otimes B_{0}+ p(1-q)A_{0}\otimes B_{1}+ (1-p)q A_{1}\otimes B_{0}-(1-p)(1-q)A_{1}\otimes B_{1}]$ being the form of CHSH-function after 
introducing bias. 

The maximum probability $P^{game}$ of winning the game is 
obtained by maximizing the function $\langle CHSH(p,q)\rangle$ for different 
theories. Such maximization was first performed in the literature for the
unbiased \cite{max_popescu} scenario and subsequently, 
for the biased case as well 
\cite{tlawson}  which we follow by treating it   in two halves of the ranges of
 the parameters $p$ and $q$. First, consider the case of $p,q\geq 1/2$.
The classical maximum is obtained using an extremal strategy where the values 
of all the observables are $+1$  giving the maximum value of the above 
CHSH-function to be 
$1-2(1-p)(1-q)$.
With this classical maximum, the winning probability is given by
\begin{equation}
P^{game}(\mathcal{S},\mathcal{T},\rho_{AB})|^{classical}_{maximum}= 1-(1-p)(1-q)
\end{equation}
This reduces to the value $\frac{3}{4}$ for the unbiased case when 
$p=q=\frac{1}{2}$.

For considering the quantum strategy, Lawson et al. \cite{tlawson} divide
the parameter space in two regions of [$p,q$] space with the first region 
corresponding to $1\geq p\geq (2q)^{-1}\geq \frac{1}{2}$ (region-1). In this region, 
\begin{equation}
\langle CHSH(p,q)\rangle \leq 1-2(1-p)(1-q)
\end{equation}
thus leading to
\begin{equation}
 P^{game}(\mathcal{S},\mathcal{T},\rho_{AB})|^{region}_1= 1-(1-p)(1-q)~~
\end{equation}
One sees that the
upper bound is the same value as that achieved by classical theory, and hence,
quantum correlation (entanglement) offers no advantage  over  classical 
correlation  in performing the specified task in this region. This result could
be restated as follows. If the bias parameters are regulated in this region, 
we can not differentiate between classical and quantum correlations using the 
upper bound of the fine grained uncertainty relation corresponding to the 
biased non-local game in context. 

Now, let us consider the other region $1\geq (2q)^{-1}> p\geq \frac{1}{2}$
(region-2), which gives the bound 
\begin{equation}
\langle CHSH(p,q)\rangle \leq \sqrt{2}\sqrt{q^{2}+(1-q)^{2}}\sqrt{p^{2}+(1-p)^{2}}.
\end{equation}
This value is greater than the classical bound. So, the regulation of the 
biasing parameters in this region discriminates among classical and quantum 
correlation. The upper bound of the fine-grained uncertainty relation (i.e., 
the maximum chance of winning the game) is in this case given by,
\begin{eqnarray}
P^{game}(\mathcal{S},\mathcal{T},\rho_{AB})&|&^{quantum}_{maximum}\nonumber\\
=\frac{1}{2}[1+\sqrt{2}\sqrt{q^{2}+(1-q)^{2}}&&\sqrt{p^{2}+(1-p)^{2}}]
\end{eqnarray}
This also reduces to the unbiased value of $[\frac{1}{2}+\frac{1}{2\sqrt{2}}]$ 
for $p=q=\frac{1}{2}$. The quantum strategy for winning this game is detailed
in ref.\cite{tlawson}. The treatments are similar for the other regions where 
both $p$ and $q$ or one of them is less than $\frac{1}{2}$, as the situations are symmetric\cite{tlawson}. On the other hand,  super-quantum correlations
in the no-signaling theory  \cite{nosig} lead to the score of the game  
\begin{eqnarray}
P^{game}|^{no-signaling}_{maximum}&=& \sum_{s,t}p(s,t)\sum_{a,b}p(a,b|s,t)\nonumber\\
= pq+ (1-p)q&+&p(1-q)+(1-p)(1-q)\nonumber\\
=1~~~~~~~~~~~~~~~~&&
\end{eqnarray}
giving the same upper bound of the fine-grained uncertainty relation in spite 
of the game being biased.

\section{A biased tripartite system}

We will now consider a biased non-local tripartite game with Alice, Bob and 
Charlie as players. Similar to the bipartite case, Alice, Bob and Charlie have 
their input binary variables (or questions) $s$, $t$ and $u$ ($s,u,t\in {0,1}$)
 corresponding to their respective two different measurements settings, and 
output binary variables (or answers) $a$, $b$ and $c$ ($a,b,c \in {0,1}$) 
corresponding to their respective outcomes of measurements. Given a rule
(i.e., the winning condition) of the game, the maximum winning probability
(having the established correspondence with the upper bound of the fine-grained
 uncertainty relation \cite{tanu}) can be calculated by considering the 
various possibilities of outcomes (along with the measurements) satisfying the 
rule. 

We consider a full-correlation box (namely, Svetlichny Box \cite{Svet}) 
for which all one and two party correlations vanish \cite{nosignal}. The game 
is won if the answers satisfy
\begin{equation}
a\oplus b\oplus c= st\oplus tu\oplus us~.
\label{rule_t}
\end{equation}
In this case, Alice intends to measure with her setting $A_{0}$ with probability
 $p$ (i.e., $p(s=0)=p$) and $A_{1}$ with probability $(1-p)$ 
(i.e., $p(s=1)=(1-p)$). Bob measures $B_{0}$ and $B_{1}$ with probabilities $q$ 
and $(1-q)$ respectively(hence, $p(t=0)=q$ and $p(t=1)=(1-q)$). Charlie 
measures with his operator $C_{0}$ with probability $r$ and $C_{1}$
with probability $(1-r)$ (therefore, $p(u=0)=r$ and $p(u=1)=(1-r)$). They share
 the state $\rho_{ABC}$, and they can communicate their measurement settings 
before the game starts.  

The winning probability is quantified as,
\begin{eqnarray}
&&P^{game}(\mathcal{S}, \mathcal{T}, \mathcal{U},\rho_{ABC})\nonumber\\
&=&\sum_{s,t,u}p(s,t,u)\sum_{a,b}p(a,b,c=x^{u}_{s,t,a,b}|s,t,u)_{\rho_{ABC}}
\label{triparty}
\end{eqnarray}
where $p(s,t,u)=p(s)p(t)p(u)$ is the probability of choosing the measurement 
settings $s$ by Alice, $t$ by Bob and $u$ by Charlie from their respective sets
 {$\mathcal{S}$}, {$\mathcal{T}$} and {$\mathcal{U}$}. $p(a,b,c|s,t,u)_{\rho_{ABC}}$ is the joint probability of getting outcomes, $a$, $b$ and $c$ for 
corresponding settings $s$, $t$ and $u$ given by,
\begin{eqnarray}
p(a,b,c&=&x^{u}_{s,t,a,b}|s,t,u)_{\rho_{ABC}}\nonumber\\
&=& \sum_{c}V(a,b,c|s,t,u)\langle A_{s}^{a} \otimes B_{t}^{b}\otimes C_{u}^{c}\rangle_{\rho_{ABC}}
\label{joint_p}
\end{eqnarray}
where $A_{s}^{a}$, $B_{t}^{b}$ and $C_{u}^{c}$ are the measurements (with the 
forms given in the treatment of bipartite system) corresponding to the setting 
$s$ and outcome $a$ at the Alice's side, setting $t$ and outcome $b$ at Bob's 
side and setting $u$ and outcome $c$ at Charlie side. $V(a,b,c|s,t,u)$ equals 
$1$ only when condition ($\ref{rule_t}$) is satisfied; otherwise, $0$. Using 
the condition ($\ref{rule_t}$) and Eq.($\ref{joint_p}$), Eq.($\ref{triparty}$) 
 simplifies to
\begin{equation}
P^{game}(\mathcal{S}, \mathcal{T}, \mathcal{U},\rho_{ABC},p,q,r)= \frac{1}{2}[1+\langle S(p,q,r)\rangle_{\rho_{ABC}}]
\end{equation}
where $S(p,q,r)$ is the Svetlichny function modified with the introduction of 
bias, given by
\begin{eqnarray}
&&S(p,q,r)~~~~~~~~~~\nonumber\\
&=& pqr A_{0}\otimes B_{0}\otimes C_{0}+ pq(1-r)A_{0}\otimes B_{0}\otimes C_{1}\nonumber\\
&&+ p(1-q)r A_{0}\otimes B_{1}\otimes C_{0}+ (1-p)qr A_{1}\otimes B_{0}\otimes C_{0}\nonumber\\
&&-p(1-q)(1-r) A_{0}\otimes B_{1}\otimes C_{1}\nonumber\\
&&- (1-p)q(1-r) A_{1}\otimes B_{0}\otimes C_{1}\nonumber\\
&&-(1-p)(1-q)r A_{1}\otimes B_{1}\otimes C_{0}\nonumber\\
&&-(1-p)(1-q)(1-r)A_{1}\otimes B_{1}\otimes C_{1}~~.
\end{eqnarray}
To find the maximum probability of winning (which is the upper bound of 
fine-grained uncertainty relation as presented in Eq.($\ref{triparty}$)), we 
need to maximize $\langle S(p,q,r)\rangle_{\rho_{ABC}}$. 

The case when all the 
three parties are quantum-correlated, has been handled only numerically in 
this context \cite{tlawson}. We will however, perform this 
maximization analytically using the scheme of bipartition modeling 
\cite{bancal}. This method is based on the fact that maximal quantum violation 
for the tripartite Svetlichny inequality has been shown \cite{bancal} even 
when the system does not feature genuine tripartite non-locality, i.e., only 
two of the three parties are correlated in a nonlocal way. Since this method
of bipartition modelling will be useful for our subsequent analysis, we
first recount here some of useful results obtained using it \cite{bancal}.
The Svetlichny 
function $S(p,q,r)$ can be rearranged as 
\begin{eqnarray}
&&S(p,q,r)\nonumber\\
&=& r[CHSH(p,q)]\otimes C_{0}+(1-r)[CHSH^{\prime}(p,q)]\otimes C_{1}\nonumber\\
&&~~~~
\label{rearrangement}
\end{eqnarray}
where,
\begin{eqnarray}
CHSH(p,q)&=&[pq A_{0}\otimes B_{0}+ p(1-q)A_{0}\otimes B_{1}\nonumber\\
~~~~+(1-p)q A_{1}&\otimes & B_{0}-(1-p)(1-q)A_{1}\otimes B_{1}]\nonumber\\
CHSH^{\prime}(p,q)&=&[pq A_{0}\otimes B_{0}- p(1-q)A_{0}\otimes B_{1}\nonumber\\
~~~~-(1-p)q A_{1}&\otimes & B_{0}-(1-p)(1-q)A_{1}\otimes B_{1}]\nonumber\\
&&~~~~~~~
\label{CHSHs}
\end{eqnarray}
Here $CHSH(p,q)$ is the traditional form of CHSH-polynomial and 
$CHSH^{\prime}(p,q)$ is an equivalent form when the mapping, $B_{0}\rightarrow B_{1}$, $B_{1}\rightarrow -B_{0}$, $q\rightarrow (1-q)$ is applied. 

Now, according to the form 
of ($\ref{rearrangement}$) let us temporarily change our point of view towards 
the game as following. The version of bipartite CHSH-game played by Alice and Bob is determined by Charlie's input setting. Assume for a moment, that when 
Charlie's input is $C_{0}$, Alice and Bob play the standard biased CHSH-game 
and when Charlie's input is $C_{1}$, they play CHSH$^{\prime}$. Alice and Bob are
 together (and separated from Charlie) and being unaware of Charlie's 
measurements, produce any bipartite non-local probability distribution. Hence 
Alice and Bob are effectively playing the average game 
$ r |\langle CHSH(p,q)\rangle|+ (1-r) |\langle CHSH^{\prime}(p,q)\rangle|$. If Alice and Bob stay separated and 
any of them be with Charlie and knows about Charlie's measurements, they will not be 
able to produce results better than the local bound. For the region 
$p,q,r \geq \frac{1}{2}$, the classical maximum is calculated to be,
\begin{equation}
\langle S\rangle_{max}=1-2(1-p)(1-q)
\end{equation}
giving
\begin{equation}
P^{game}(\mathcal{S}, \mathcal{T}, \mathcal{U},\rho_{ABC})|^{classical}_{maximum}= 1-(1-p)(1-q)
\end{equation}
which reduces to the value $\frac{3}{4}$ for the unbiased game ($r$ is averaged
 off due to both the Bell functions possessing the same classical maximum). 

In order to treat the quantum optimization we consider that the 
three parties share 
a three-qubit Greenberger-Horne-Zeilinger (GHZ) state $|\psi\rangle= \frac{1}{\sqrt{2}}|000\rangle +|111\rangle$ (for the unbiased case, the maximum violation 
of the Svetlichny function occurs for the GHZ state\cite{S1GHZ}). In this process, generally, Charlie 
needs to choose two measurements in a way that he prepares two qubit entangled states(for Alice and Bob) which will maximize their corresponding CHSH-functions simultaneously. The purpose of this strategic choice of measurements by Charlie is to maximize the Svetlichny function and hence to improve the score of the non-local game to its best. Consider the choice being, 
$C_{0}=\sigma_{x}$ and $C_{1}=-\sigma_{y}$ which prepare the states 
$|\phi_{\pm}\rangle= \frac{1}{\sqrt{2}}(|00\rangle \pm|11\rangle)$ and 
$|\tilde{\phi}_{\pm}\rangle= \frac{1}{\sqrt{2}}(|00\rangle\pm i|11\rangle)$ 
respectively, for Alice and Bob. Note that
\begin{equation}
(I\otimes U_{B})\tilde{\rho_{\pm}}(I\otimes U_{B}^{\dagger})= \rho_{\pm}
\end{equation}
where $\tilde{\rho_{\pm}}=|\tilde{\phi}_{\pm}\rangle\langle \tilde{\phi}_{\pm}|$ and $\rho_{\pm}=|\phi_{\pm}\rangle\langle\phi_{\pm}|$ and $U_{B}$ is a unitary rotation on the Bob's qubit, given by
\begin{equation}
U_{B}=\left( \begin{array}{cc}
1 & 0 \\ 
0 & -i
\end{array} \right)
\end{equation}
and consequently,  the equivalence of optimizations of $CHSH(p,q)$ and 
$CHSH^{\prime}(p,q)$ is realized as,
\begin{eqnarray}
&&|\langle\phi_{\pm}| CHSH(p,q)|\phi_{\pm}\rangle |\nonumber\\
&&= |\langle\tilde{\phi_{\pm}}|(I\otimes U^{\dagger}_B)CHSH^{\prime}(p,q)(I\otimes U_{B})|\tilde{\phi_{\pm}}\rangle |
\end{eqnarray}
or simply,
\begin{equation}
\langle CHSH(p,q)\rangle_{\rho_{\pm}}=\langle CHSH^{\prime}(p,q)\rangle_{\rho_{\pm}}
\end{equation}
The above equation is true provided the aforesaid mapping between the operators $B_{0}$ and $B_{1}$ and their probability distribution $q$ (i.e., the mapping $B_{0}\rightarrow B_{1}$, $B_{1}\rightarrow -B_{0}$, $q\rightarrow (1-q)$) is considered. So, as we focus on achieving the best score for the present nonlocal game, we may now 
think of the situation (instead of Alice and Bob playing with two kinds of CHSH games) as only the standard CHSH-game  
being played that is averaged over the scenarios when 
Bob rotates unitarily his qubit before measurement and when he does not. The 
unitary rotation preserves the nonlocal property of the state causing no 
discrepancy. In the region $p,q,r\geq \frac{1}{2}$, the maximum value of 
$\langle S(p,q,r)\rangle$ is calculated (using a procedure for maximizing 
$CHSH(p,q)$ similar to the bipartite case) to be,
\begin{equation}
\langle S(p,q,r)\rangle_{|GHZ\rangle}|_{max}^{1}= 1-2(1-p)(1-q)
\label{quantum1}
\end{equation}
for the region $1\geq p\geq (2q)^{-1}\geq \frac{1}{2}$ which is the same as the
 classically achieved upper bound. Here $\langle S\rangle$ is not a function of $r$ because 
the nonlocal strength of Alice's and Bob's systems are identical for the two 
different measurements of Charlie. For the 
region $1\geq (2q)^{-1}> p\geq \frac{1}{2}$, one obtains
\begin{equation}
\langle S(p,q,r)\rangle_{|GHZ\rangle}|_{max}^{2}= \sqrt{2}\sqrt{q^{2}+(1-q)^{2}}\sqrt{p^{2}+(1-p)^{2}}
\label{quantum2}
\end{equation}
The bound ($\ref{quantum2}$) is greater than the bound ($\ref{quantum1}$), and hence, the quantum correlation 
dominates here.  The expression for maximum winning probability in this case
is given by
\begin{eqnarray}
&&P^{game}(\mathcal{S}, \mathcal{T}, \mathcal{U},\rho_{ABC})|^{quantum}_{maximum}\nonumber\\
&=&\frac{1}{2}[1+\sqrt{2}\sqrt{q^{2}+(1-q)^{2}}\sqrt{p^{2}+(1-p)^{2}}]
\end{eqnarray}
For every $r$($\neq 0,1$) there is the same patch in the $p-q$ 
space which separates the classical and quantum correlations in terms of their
degree of non-locality. 

It may be noted that 
the results for the tripartite system is  quantitatively somewhat different from
 the numerical calculation provided by Lawson ${\it et. al.}$ \cite{tlawson}. 
According to the latter if all the biasing parameters $(p, q, r)$ are made 
equal, the no-quantum-advantage region is above 
$p\simeq 0.8406$ which is slightly different from  $p\simeq 0.7071$ for our 
case. This deviation reflects the fact that the use of the bipartition model 
\cite{bancal} does not, in general, capture all types of tripartite nonlocal 
correlations.
Finally, for no-signaling theory the upper bound turns out to be $1$, as 
expected. Note also, that in the other regions when all $p$, $q$ and $r$ or one or two of them are less than $\frac{1}{2}$, the 
treatments are similar, as in the bipartite case.

\section{Conclusions}

In this work we have employed the  fine-grained uncertainty relation 
\cite{FUR1} 
to distinguish between 
classical, quantum and super-quantum correlations based on their strength of
nonlocality, in the context of biased games \cite{tlawson}
involving two or three parties. Discrimination among the underlying theories 
with different degrees of nonlocality is possible for a
particular range of the biasing parameters. This range of bias parameters turns out to
be the region for which quantum correlations offer the
advantage of winning the said nonlocal game over classical correlations.
For the tripartite game
in case of no bias, the Svetlichny inequality is able to discriminate \cite{tanu}
 among classical, quantum and super-quantum correlations. But in the presence
of bias, using a bipartition model \cite{bancal} we observe here that there 
is a zone specified by the biasing 
parameters where even the Svetlichny inequality cannot perform 
this discrimination. 
The  extent of non-locality that can be captured by the fine-grained uncertainty
principle thus turns out to be regulated by the bias parameters.  Our approach, 
in spite of featuring a narrower range of the biasing parameters providing 
quantum advantage, serves the purpose of developing an analytical approach 
to explore the connection between biased nonlocal retrieval games and 
the upper bound of fine-grained uncertainty capturing the nonlocal strengths
of various correlations. Analytical generalizations to multiparty nonlocal 
games may indeed be feasible using this approach.

{\it Acknowledgments:} ASM acknowledges
support from the DST project no. SR/S2/PU-16/2007. TP thanks UGC, India for financial support.

\end{document}